\documentclass[]{spie}  

\usepackage{amsmath,amsfonts,amssymb}
\usepackage{graphicx}
\usepackage[colorlinks=true, allcolors=blue]{hyperref}
\usepackage[leftcaption]{sidecap}

\title{Large-scale Augmented Granger Causality (lsAGC) for Connectivity Analysis in Complex Systems: From Computer Simulations to Functional MRI (fMRI)}

\author[a,b,c,d]{Axel Wismüller,}
\author[a]{M. Ali Vosoughi}

\affil[a]{Department of Electrical and Computer Engineering, University of Rochester, NY, USA}
\affil[b]{ Department of Imaging Sciences, University of Rochester, NY, USA}
\affil[c]{Department of Biomedical Engineering, University of Rochester, NY, USA}
\affil[d]{Faculty of Medicine and Institute of Clinical Radiology, Ludwig Maximilian University,
Munich, Germany}

\authorinfo{Further author information: (Send correspondence to Ali Vosoughi)\\Ali Vosoughi: E-mail: mvosough@ur.rochester.edu}
\begin{document} 
\maketitle
\begin{abstract}

We introduce large-scale Augmented Granger Causality (lsAGC) as a method for connectivity analysis in complex systems. The lsAGC algorithm combines dimension reduction with source time-series augmentation and uses predictive time-series modeling for estimating directed causal relationships among time-series. This method is a multivariate approach, since it is capable of identifying the influence of each time-series on any other time-series in the presence of all other time-series of the underlying dynamic system. We quantitatively evaluate the performance of lsAGC on synthetic directional time-series networks with known ground truth. As a reference method, we compare our results with cross-correlation, which is typically used as a standard measure of connectivity in the functional MRI (fMRI) literature. Using extensive simulations for a wide range of time-series lengths and two different signal-to-noise ratios of 5 and 15 dB,  lsAGC consistently outperforms cross-correlation at accurately detecting network connections, using Receiver Operator Characteristic Curve (ROC) analysis, across all tested time-series lengths and noise levels. In addition, as an outlook to possible clinical application, we perform a preliminary qualitative analysis of connectivity matrices for fMRI data of Autism Spectrum Disorder (ASD) patients and typical controls, using a subset of 59 subjects of the Autism Brain Imaging Data Exchange II (ABIDE II) data repository. Our results suggest that lsAGC, by extracting sparse connectivity matrices, may be useful for network analysis in complex systems, and may be applicable to clinical fMRI analysis in future research, such as targeting disease-related classification or regression tasks on clinical data.
\end{abstract}

\section{INTRODUCTION} \label{sec:intro}  

Currently, the quantification of directed information transfer between interacting brain areas is one of the most challenging methodological problems in computational neuroscience. A fundamental problem is identifying connectivity in very high-dimensional systems. A common practice has been to transform a high-dimensional system into a simplified representation, e.g. by clustering, Principal, or Independent Component Analysis. The drawback of such methodology is that an identified interaction between such simplified components cannot readily be transferred back into the original high-dimensional space. Thus, directed interactions between the original network nodes can no longer be revealed. Although this significantly limits the interpretation of brain network activities in physiological and disease states, surprisingly little effort has been devoted to circumvent the inevitable information loss induced by the aforementioned frequently employed techniques.

Various methods have been proposed to obtain directional relationships in multivariate time-series data, e.g., transfer entropy [\citeonline{schreiber2000measuring}] and mutual information [\citeonline{kraskov2004estimating}]. However, as the multivariate problem's dimensions increase, computation of the density function becomes computationally expensive [\citeonline{mozaffari2019online,mozaffari2019online_ieee}]. Under the Gaussian assumption, transfer entropy is equivalent to Granger causality [\citeonline{barnett2009granger}]. However, the computation of multivariate Granger causality for short time series in large-scale problems is challenging [\citeonline{vosoughi2020large,dsouza2020large}]. To address these problems, we have previously proposed a method for multivariate Granger causality analysis using linear multivariate auto-regressive (MVAR) modeling, which simultaneously circumvents the drawbacks of above mentioned simplification strategies by introducing an invertible dimension reduction followed by a back-projection of prediction residuals into the original data space (large-scale Granger Causality, lsGC) [\citeonline{d2016large_66xx}]. We have also demonstrated the applicability of this approach to resting-state fMRI analysis [\citeonline{61_dsouza2017exploring}]. Recently, we have also presented an alternative multivariate Granger causality analysis method, large-scale Extended Granger Causality (lsXGC), that uses an augmented dimension-reduced time-series representation for predicting target time-series in the original high-dimensional system directly, i.e., without inverting the dimensionality reduction step [\citeonline{vosoughi2021marijuana}].

In this paper, we introduce a hybrid of both methods, large-scale Augmented Granger Causality (lsAGC) that combines both invertible dimension reduction and time-series augmentation. It first uses an augmented dimension-reduced time-series representation for prediction in the low-dimensional space, followed by an inversion of the initial dimension reduction step. In the following, we explain the lsAGC algorithm and present quantitative results on synthetic time-series data with known connectivity ground truth. Finally, as an outlook to possible clinical application, we perform a preliminary qualitative analysis of connectivity matrices for fMRI data of Autism Spectrum Disorder (ASD) patients and typical controls, using a subset of the Autism Brain Imaging Data Exchange II (ABIDE II) data repository. \newline
This work is embedded in our group’s endeavor to expedite artificial intelligence in biomedical imaging by means of advanced pattern recognition and machine learning methods for computational radiology and radiomics, e.g., [ \citeonline{9_leinsinger2006cluster,10_wismuller2004fully,nattkemper2005tumor,bunte2010adaptive,8_wismueller2000segmentation,11_hoole2000analysis,12_wismuller2006exploratory,13_wismuller1998neural,14_wismuller2002deformable,15_behrends2003segmentation,16_wismuller1997neural,17_bunte2010exploratory,18_wismuller1998deformable,19_wismuller2009exploration,20_wismuller2009method,22_huber2010classification,23_wismuller2009exploration,24_bunte2011neighbor,25_meyer2004model,26_wismuller2009computational,27_meyer2003topographic,28_meyer2009small,29_wismueller2010model,meyer2007unsupervised,30_huber2011performance,31_wismuller2010recent,meyer2007analysis,32_wismueller2008human,wismuller2015method,33_huber2012texture,34_wismuller2005cluster,35_twellmann2004detection,37_otto2003model,38_varini2004breast,39_huber2011prediction,40_meyer2004stability,41_meyer2008computer,42_wismuller2006segmentation,45_bhole20143d,46_nagarajan2013computer,47_wismuller2004model,48_meyer2004computer,49_nagarajan2014computer,50_nagarajan2014classification,yang2014improving,wismuller2014pair,51_wismuller2014framework,schmidt2014impact,wismuller2015nonlinear,wismuller2016mutual,52_schmidt2016multivariate,abidin2017using,53_chen2018mri,54_abidin2018alteration,55_abidin2018deep,dsouza2018mutual,chockanathan2019automated, vosoughi2021schizophrenia} ].


\section{DATA} \label{sec:data}

\subsection{Synthetic Networks with Known Ground Truth for Quantitative Analysis}

We quantitatively evaluate the performance of lsAGC for network structure recovery using synthetic networks with known ground truth. We constructed ground truth networks with N = 50 nodes, each containing 5 modules of 10 nodes with high (low) probability for the existence of directed intra- (inter-) module connections. We simulated two values of additive white Gaussian noise, with signal-to-noise ratios (SNR) of 15 dB and 5 dB, and repeated the experiment 100 times with different noise seed. Networks were realized as noisy stationary multivariate auto-regressive (MVAR) processes of model order p = 2 in each of T = 1000 temporal process iterations. The network structure was adapted from [\citeonline{baccala2001partial}] and [\citeonline{wismuller2014pair}]. 

\subsection{Functional MRI Data for Qualitative Analysis}

The following explanation of participants and data in this section follows the description in [\citeonline{56_dsouza2019classification}]: The
Autism Brain Imaging Data Exchange II (ABIDE II) initiative has made publicly available MRI data from
multiple sites. We used resting-state fMRI data from 59 participants of the online ABIDE II repository (\hyperlink{http://
fcon_1000.projects.nitrc.org/indi/abide}{http://
fcon\_1000.projects.nitrc.org/indi/abide}) in this analysis, namely the data from the Olin Neuropsychiatry
Research Center, Institute of Living at Hartford Hospital [\citeonline{57_first1995structured}].

This data set consists of 24 ASD subjects (18-31 years) and 35 typical controls (18-31 years). Autism diagnosis
was based on the ASD cutoff of the Autism Diagnostic Observation Schedule-Generic (ADOS-G). The typical
controls (TC) were screened for autism using the ADOS-G and any psychiatric disorder based on the Structured
Clinical Interview for DSM-IV Axis I Disorders-Research Version (SCID-I RV) [\citeonline{57_first1995structured}].

In this data set, resting-state fMRI scans were obtained from all subjects using Siemens Magnetom Skyra Syngo MR D13. The study protocol
included: (i) High-resolution structural imaging using T1-weighted magnetization-prepared rapid gradient-echo (MPRAGE) sequence, TE = 2.88 ms, TR = 2200 ms, isotropic voxel size 1 mm, flip angle $13^{\circ}$. (ii) Resting-state fMRI scans with TE = 30 ms, TR = 475 ms, flip angle $60^{\circ}$, isotropic voxel size of 3 mm. The acquisition lasted 7 minutes and 37 seconds.

The fMRI data used in this study were pre-processed using standard methodology. Motion correction,
brain extraction, and correction for slice timing acquisition were performed. Additional nuisance regression for removing variations due to head motion and physiological processes was carried out. Each data set was finally registered to the 2 mm MNI standard space using a 12-parameter affine transformation. The functional data used in this study was pre-processed using the CONN toolbox (\hyperlink{www.nitrc.org/projects/conn}{www.nitrc.org/projects/conn})  [\citeonline{58_whitfield2012conn}]. Additionally, the time-series were normalized to zero mean and unit standard deviation to focus on signal dynamics rather than amplitude  [\citeonline{59_wismuller2002cluster}]. Finally, the brain was parcellated into 90 regions as defined by the Automated Anatomic Labelling (AAL) template [\citeonline{60_tzourio2002automated}]. Each regional time-series was represented by the average time-series of all voxels included in the region.

\section{ALGORITHM}\label{sec:methods}
Large-scale Augmented Granger Causality (lsAGC) has been developed based on 1) the principle of original Granger
causality, which quantifies the causal influence of time-series $\mathbf{x_s}$ on time-series $\mathbf{x_t}$ by quantifying the amount of improvement in the prediction of $\mathbf{x_t}$ in presence of $\mathbf{x_s}$. 2) the idea of dimension reduction, which resolves the problem of the tackling a under-determined system, which is frequently faced in fMRI analysis, since the number of acquired temporal samples usually is not sufficient for estimating the model parameters \cite{61_dsouza2017exploring}.

Consider the ensemble of time-series $\mathcal{X}\in \mathbb{R}^{N\times T}$, where $N$ is the number of time-series (Regions Of Interest – ROIs) and $T$ the number of temporal samples. Let $\mathcal{X} = (\mathbf{x_1}, \mathbf{x_2}, \dots, \mathbf{x_N})^{\mathsf{T}}$ be the whole multidimensional system and $x_i \in \mathbb{R}^{1\times T}$ a single time-series with $i = 1, 2, \dots,N$, where $\mathbf{x_i} = (x_i(1), x_i(2), \dots, x_i(T))$. In order to overcome the under-determined problem, first $\mathcal{X}$ will be decomposed into its first $p$ high-variance principal components
$\mathcal{Z} \in \mathbb{R}^{p\times T}$ using Principal Component Analysis (PCA), i.e.,
\begin{equation}
\mathcal{Z}=W\mathcal{X},    
\end{equation}
where $W\in \mathbb{R}^{p\times N}$ represents the PCA coefficient matrix. Subsequently, the dimension-reduced time-series ensemble $\mathcal{Z}$ is augmented by one original time-series $\mathbf{x_s}$ yielding a dimension-reduced augmented time-series ensemble $\mathcal{Y}\in \mathbb{R}^{(p+1)\times T}$ for estimating the influence of $\mathbf{x_s}$ on all other time-series.

Following this, we locally predict the dimension-reduced representation $\mathcal{Z}$ of the original high-dimensional system $\mathcal{X}$ at each time sample $t$, i.e. $\mathcal{Z}(t)\in \mathbb{R}^{p\times 1}$ by calculating an estimate $\hat{\mathcal{Z}}_{\mathbf{x_s}}(t)$. To this end, we fit an affine model based on a vector of $m$ vector of m time samples of $\mathcal{Y}(\tau)\in \mathbb{R}^{(p+1)\times 1}$($\tau=t-1, t-2, \dots, t-m$), which is $\mathbf{y}(t)\in \mathbb{R}^{m.(p+1)\times 1}$, and a parameter matrix $\mathcal{A}\in \mathbb{R}^{p\times m.(p+1)}$ and a constant bias vector $\mathbf{b}\in \mathbb{R}^{p\times 1}$, 
\begin{equation}\label{eqn:low_dim_estimate}
    \hat{\mathcal{Z}}_{\mathbf{x_s}}(t)=\mathcal{A}\mathbf{y}(t)+\mathbf{b},~~ t=m+1, m+2, \dots, T.
\end{equation}

Subsequently, we use the prediction $\hat{\mathcal{Z}}_{\mathbf{x_s}}(t)$ to calculate an estimate of $\mathcal{X}$ at time $t$, i.e. $\mathcal{X}(t)\in \mathbb{R}^{N\times 1}$ by inverting the PCA of equation \eqref{eqn:low_dim_estimate}, i.e.
\begin{equation}
    \mathcal{X}=W^{\dagger}\mathcal{Z},
\end{equation}
where $W^{\dagger}\in \mathbb{R}^{N\times p}$ represents the inverse of the PCA coefficient matrix $W$, which is calculated as the \textit{Moore–Penrose} pseudoinverse of $W$.

Now $\hat{\mathcal{X}}_{\setminus {\mathbf{x_s}}}(t)$, which is the prediction of $\mathcal{X}(t)$ without the information of $\mathbf{x_s}$, will be estimated. The estimation processes is identical to the previous one, with the only difference being that we have to remove the augmented time-series $\mathbf{x_s}$ and its corresponding column in the PCA coefficient matrix $W$.

The computation of a lsAGC index is based on comparing the variance of the prediction errors obtained with
and without consideration of $\mathbf{x_s}$. The lsAGC index $f_{\mathbf{x_s}\xrightarrow{}\mathbf{x_t}}$ , which indicates the influence of $\mathbf{x_s}$ on $\mathbf{x_t}$, can be calculated by the following equation:
\begin{equation}
    f_{\mathbf{x_s}\xrightarrow{}\mathbf{x_t}}=\log {\frac{\mathrm{var}(e_s)}{\mathrm{var}(e_{\setminus s})}},
\end{equation}
where $e_{\setminus s}$ is the error in predicting $\mathbf{x_t}$ when $\mathbf{x_s}$ was not considered, and $e_s$ is the error, when $\mathbf{x_s}$ was used. Based on preliminary analyses, in this study, we set $p = 7$ and $m = 3$.

\section{RESULTS}\label{sec:results}
\begin{figure}
         \includegraphics[width=\textwidth]{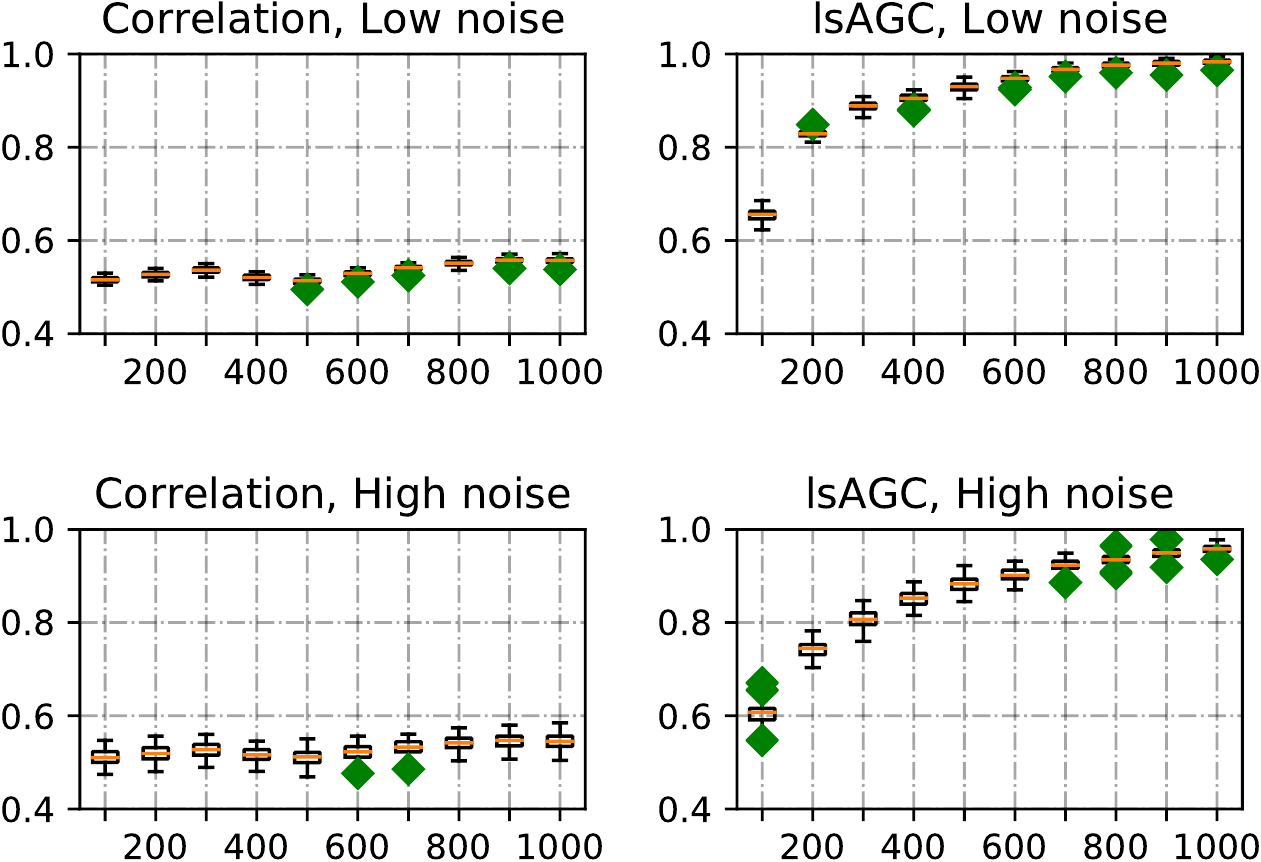}
        \caption{Quantitative performance comparison of cross-correlation and lsAGC for recovery of synthetic networks. The vertical axis is the Area Under the Curve (AUC) for Receiver Operating Characteristics (ROC) analysis, where an AUC = 1 indicates a perfect network recovery and AUC = 0.5 random assignment. Whiskers are related to the 95\% confidence interval, green diamonds represent outliers, orange lines represent medians, and boxes are drawn from the first quartile to the third quartile. It is clearly seen that lsAGC outperforms cross-correlation over all tested time-series lengths and noise levels.}
        \label{fig:auc_plots}
\end{figure}

\textbf{Quantitative Analysis of Synthetic Networks with Known Ground Truth:} Network reconstruction results for the synthetic networks with known ground truth, using the Area Under the Curve (AUC) for Receiver Operating Characteristic (ROC) analysis, are shown in Fig. \ref{fig:auc_plots}. For each time-series length and each noise level, we performed 100 simulations.
As can be seen from Fig. \ref{fig:auc_plots}, lsAGC consistently outperforms cross-correlation in its ability to accurately recover network structure over a wide range of time-series-lengths in both high- and low-noise scenarios, with a mean AUC for lsAGC for a time-series length of 1000 temporal samples equal to 98.9\% and 97.1\% for signal-to-noise values of 15 and 5 dB, respectively. On the other hand, cross-correlation performs quite poorly compared to lsAGC with its mean AUC ranging around 0.5 for all examined time-series lengths and noise levels, equivalent to the quality of randomly guessing the presence or absence of network connections.  

\begin{figure}
    \includegraphics[width=\textwidth]{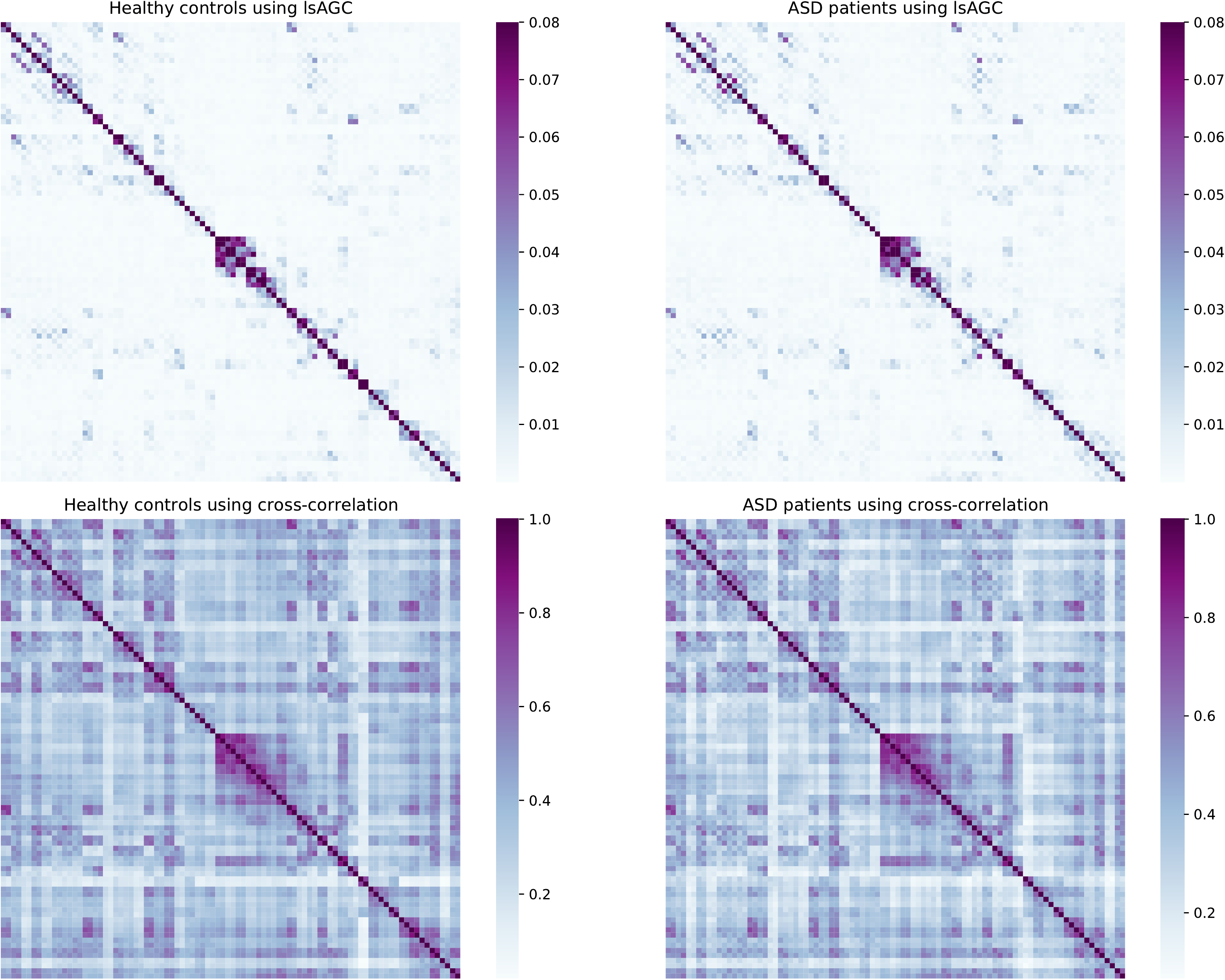}
    \caption{Averaged connectivity matrices: top left: average connectivity matrix of healthy control subjects using lsAGC, top right: average connectivity matrix of ASD patients using lsAGC, bottom left: average connectivity matrix of healthy control subjects using cross-correlation, and bottom right: average connectivity matrix of ASD patients using cross-correlation. Note that the different methods capture different connectivity features, and that there are slight differences of connectivity patterns between healthy subjects and ASD patients. Also, the lsAGC connectivity matrices appear to be significantly “sparser” than cross-correlation matrices. This observation is quantitatively confirmed by calculating the entropy over the matrix elements, as explained in the text. }
    \label{fig:matrix_plot}
\end{figure}

\textbf{Qualitative Analysis of Connectivity Matrices Extracted from fMRI Data: }
Averaged connectivity matrices, which were extracted using lsAGC and cross-correlation, are shown in Fig. \ref{fig:matrix_plot} for both healthy controls and ASD patients. These matrices were obtained by calculating and then averaging over the connectivity matrices of the 24 ASD patients and 35 typical controls, using the proposed lsAGC algorithm as well as conventional cross-correlation analysis. Visual inspection of the mean connectivity matrices in Fig. \ref{fig:matrix_plot} reveals subtle differences between ASD patients and healthy controls for both methods, which may be exploited for classification among the two cohorts in future research. We also find from visual inspection of Fig. \ref{fig:matrix_plot} that the features extracted by the two methods are likely different, where the mean connectivity matrices for lsAGC appear to be more “sparse” than for cross-correlation. To quantify this qualitative visual impression, we calculated the entropy of the connectivity matrix elements for each of the 59 subjects as a surrogate for matrix “sparseness”.  The mean entropy for healthy controls with lsAGC and correlation was 1.66 $\pm$ 0.14 and 4.76 $\pm$ 0.07, respectively, and the mean entropy for the ASD patients with lsAGC and correlation was 1.75 $\pm$ 0.12 and 4.78 $\pm$ 0.07, respectively. I.e., for both cohorts, the sparseness of lsAGC connectivity matrices appears to be higher than for cross-correlation analysis. We found that this difference between methods, as expressed by the entropy of the connectivity matrix elements, was statistically significant (Mann-Whitney U-test, $p < 10^{-8}$). We conclude that lsAGC may be useful for disease-related classification or regression tasks on clinical fMRI data, because it may extract relevant features potentially not captured by cross-correlation, which is currently used as the mainstay of fMRI connectivity analysis. This hypothesis can be further investigated in future research.

\section{CONCLUSIONS}\label{sec:conclusions}

In this work, we have introduced large-scale Augmented Granger Causality (lsAGC) as a method for connectivity analysis in complex systems. The lsAGC algorithm combines dimension reduction with source time-series augmentation and uses multivariate predictive time-series modeling for estimating directed causal relationships among time-series. We quantitatively evaluated the performance of lsAGC on synthetic directional time-series networks with known ground truth. Using simulations for a wide range of time-series lengths and different signal-to-noise ratios, we compared lsAGC with cross-correlation, which is currently used as the clinical standard for fMRI connectivity analysis. We found that lsAGC consistently outperformed cross-correlation at accurately detecting network connections. In addition, we performed a preliminary qualitative analysis of connectivity matrices for fMRI data of Autism Spectrum Disorder (ASD) patients and typical controls, using a subset of the ABIDE II data repository. Our results suggest that lsAGC, by extracting sparse connectivity matrices, may be useful for network analysis in complex systems, and may be applicable to clinical fMRI analysis in future research, such as targeting disease-related classification or regression tasks on clinical data.

\acknowledgments 
 
This research was funded by Ernest J. Del Monte Institute for Neuroscience Award from the Harry T. Mangurian Jr. Foundation. This work was conducted as a Practice Quality Improvement (PQI) project related to American Board of Radiology (ABR) Maintenance of Certificate (MOC) for Prof. Dr. Axel Wismüller. This work is not being and has not been submitted for publication or presentation elsewhere.  

\bibliography{report} 
\bibliographystyle{spiebib} 

\end{document}